# Struggles with Survey Weighting and Regression Modeling[1]

**Andrew Gelman**

*Abstract.* The general principles of Bayesian data analysis imply that models for survey responses should be constructed conditional on all variables that affect the probability of inclusion and nonresponse, which are also the variables used in survey weighting and clustering. However, such models can quickly become very complicated, with potentially thousands of poststratification cells. It is then a challenge to develop general families of multilevel probability models that yield reasonable Bayesian inferences. We discuss in the context of several ongoing public health and social surveys. This work is currently open-ended, and we conclude with thoughts on how research could proceed to solve these problems.

*Key words and phrases:* Multilevel modeling, poststratification, sampling weights, shrinkage.

## 1. BACKGROUND

Survey weighting is a mess. It is not always clear how to use weights in estimating anything more complicated than a simple mean or ratios, and standard errors are tricky even with simple weighted means. (Software packages such as Stata and SUDAAN perform analysis of weighted survey data, but it is not always clear which, if any, of the available procedures are appropriate for complex adjustment schemes. In addition, the construction of weights is itself an uncodified process.) Contrary to what is assumed by many theoretical statisticians, survey weights are *not* in general equal to inverse probabilities of selection but rather are typically constructed based on a combination of probability calculations and nonresponse adjustments.

Regression modeling is a potentially attractive alternative to weighting. In practice, however, the potential for large numbers of interactions can make regression adjustments highly variable. This paper reviews the motivation for hierarchical regression, combined with poststratification, as a strategy for correcting for differences between sample and population. We sketch some directions toward a practical solution, which unfortunately has not yet been reached.

### 1.1 Estimating Population Quantities from a Sample

Our goal is to use sample survey data to estimate a population average or the coefficients of a regression model. The regression framework also includes small-area estimation, since that is simply a regression on a discrete variable corresponding to indicators for the small areas.

We shall consider two running examples: a series of CBS/New York Times national polls from the 1988 election campaign, and the New York City Social Indicators Survey, a biennial survey of families that was conducted by Columbia University's School of Social Work (Garfinkel and Meyers, 1999; Meyers and Teitler, 2001; Garfinkel et al., 2003). Both sets of surveys used random digit dialing.

*Andrew Gelman is Professor of Statistics and Professor of Political Science, Department of Statistics, Columbia University, New York, New York 10027, USA e-mail: gelman@stat.columbia.edu.*

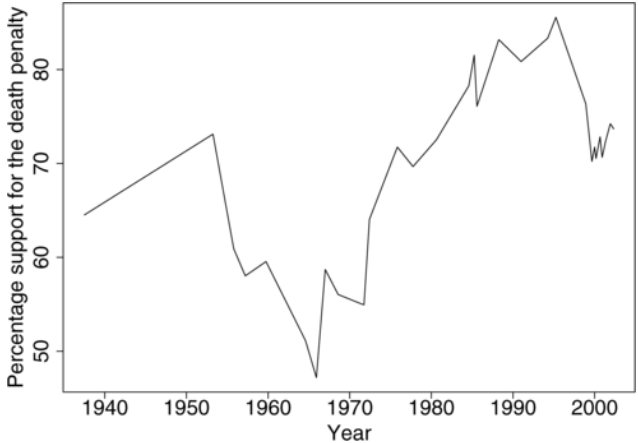

Fig. 1. *The proportion of adults surveyed who answered yes in the Gallup Poll to the question, "Are you in favor of the death penalty for a person convicted of murder?" among those who expressed an opinion on the question. It would be interesting to estimate these trends in individual states.*

For the pre-election polls, our quantity of primary interest is the proportion of people who support the Republican candidate for President in the country or in each state [or the proportion of *voters* who support the Republican candidate, which is a ratio: the proportion of people who will vote and support the Republican, divided by the proportion who will support the Republican; it is straightforward to move from estimating a population mean to estimating this ratio, as discussed in the context of this example by Park, Gelman and Bafumi (2004)]. We would also like to use series of national polls to estimate state-by-state time trends, for example in the support for the death penalty over the past few decades. (See Figure 1 for the national trends.)

For the Social Indicators Survey, we are interested in population average responses to questions such as, "Do you rate the schools as poor, fair, good or very good?", average responses in subpopulations (e.g., the view of the schools among parents of school-age children), and so-called "analytical" studies that can be expressed in terms of regressions (e.g., predicting total satisfaction given demographics and specific attitudes about health care, safety, etc.). In this article, we focus on trends from 1999 to 2001, as measured by changes in two successive Social Indicators Surveys, on a somewhat arbitrary selection of questions chosen to illustrate the general concerns of the survey.

Table 1 shows the questions, the estimated average responses in each year, and the estimated differences and standard errors as obtained using two different methods of inference. This paper is centered on the puzzle of how these two estimation methods differ. We shall get back to this question in a moment after reviewing some basic ideas in survey sampling inference.

### 1.2 Poststratification and Weighting

Naive promulgators of Bayesian inference—or the modeling approach to inference in general—used to say that the method of data collection was irrelevant to estimation from survey data. All that matters, from this slightly misguided perspective, is the likelihood, or the model of how the data came to be. However, as has been pointed out by Rubin (1976), the usual Bayesian or likelihood analysis implicitly assumes the design is "ignorable," which in a sampling context roughly means that the analysis includes all variables that affect the probability of a

TABLE 1

| Question | Weighted averages | | (a) time change in percent | (b) linear regression coefficient of time | (a) time change on logit scale | (b) logistic regression coefficient of time |
|---|---|---|---|---|---|---|
| | 1999 | 2001 | | | | |
| Adult in good/excellent health | 75% | 78% | 3.4% (2.4%) | 6.6% (1.4%) | 0.19 (0.13) | 0.48 (0.10) |
| Child in good/excellent health | 82% | 84% | 1.7% (1.5%) | 1.2% (1.3%) | 0.24 (0.21) | 0.18 (0.20) |
| Neighborhood is safe/very safe | 77% | 81% | 4.5% (2.3%) | 4.1% (1.5%) | 0.27 (0.14) | 0.27 (0.10) |

Estimates for some responses from two consecutive waves of the New York City Social Indicators Survey, and estimated changes, with standard errors in parentheses. Changes are estimated in percentages and on the logit scale. In each scale, two estimates are presented: (a) simple differences in weighted means and (b) regression controlling for the variables used in the weighting. Approaches (a) and (b) can give similar results but sometimes are much different.

person being included in the survey (see Chapter 7 of Gelman et al., 2004, for a review).

In a regression context, the analysis should include, as "$X$ variables," everything that affects sample selection or nonresponse. Or, to be realistic, all variables should be included that have an important effect on sampling or nonresponse, if they also are potentially predictive of the outcome of interest. In a public survey such as the CBS polls, a good starting point is the set of variables used in their weighting scheme: number of adults and number of telephone lines in the sampled household; region of the country; and sex, ethnicity, age and education level of the respondent (see Voss, Gelman and King, 1995). For the Social Indicators Survey, we did our own weighting (Becker, 1998) using similar information: number of telephone lines (counted as 1/2 for families with intermittent phone service), number of adults and children in the family, and ethnicity, age and education of the head of household. Weights for each survey are constructed by multiplying a series of factors.

In the sampling context, ignorability corresponds to the assumption of simple random sampling within poststratification cells or, more generally, the assumption that, within poststratification cells, the relative probabilities of selection are equal. (This is the information used in constructing sampling weights.) Adjustment for unit nonresponse is implicit in this framework; for example, by poststratifying on sex, an analysis adjusts simultaneously for differences between men and women in probability of inclusion in the sample (i.e., probability of being sampled, multiplied by probability of responding). We shall ignore item nonresponse (or, equivalently, suppose any missing data have been randomly imputed; see the discussion in Rubin, 1996).

We now review the unified notation for poststratification and survey weighting of Little (1991, 1993) and Gelman and Carlin (2002); see also Holt and Smith (1979). Here we use the notation $y, z$ for variables that are observed in the sample only, and $X$ for variables that are observed in the sample and known in the population. For simplicity, we assume throughout this article that the population size is large, so that the finite-population quantities of interest (averages, population totals or regression coefficients) are essentially the same as the corresponding superpopulation quantities.

*Poststratification.* The purpose of poststratification is to correct for known differences between sample and population. In the basic formulation, we have variables $X$ whose joint distribution in the population is known, and an outcome $y$ whose population distribution we are interested in estimating. We shall assume $X$ is discrete, and label the possible categories of $X$ as *poststratification cells $j$*, with population sizes $N_j$ and sample sizes $n_j$. In this notation, the total population size is $N = \sum_{j=1}^{J} N_j$ and the sample size is $n = \sum_{j=1}^{J} n_j$. The implicit model of poststratification is that the data are collected by simple random sample within each of the $J$ poststrata. The assignment of sample sizes to poststrata is irrelevant. In fact, classical stratification (in which the sampling really *is* performed within strata) is a special case of poststratification as we formulate it. We assume the population size $N_j$ of each category $j$ is known. These categories include all the cross-classifications of the predictors $X$. [In some cases the cell populations are unknown and must be estimated. For example, in the Social Indicators Survey, we adjust to estimated demographics from the Current Population Survey, which includes about 2000 New York City residents each year. This is enough to give reliable estimates of one-way and two-way margins (e.g., the proportion of city residents who are white females, white males, black females, black males, etc.), but the counts are too sparse to directly estimate deep interactions (e.g., the proportion who are white females, 30–45, married, with less than a high school education, etc.). The usual practical solution in this case is to poststratify on the margins (e.g., raking; see, e.g., Deville, Sarndal and Sautory, 1993). If the whole table of population counts is required, it can be estimated using iterative proportional fitting (Deming and Stephan, 1940) which sets interactions to be as small as possible while being consistent with the available population data. For this paper, we shall ignore this difficulty and treat the full vector of $N_j$'s as known.]

The population mean of any survey response can be written as a sum over poststrata,

(1) $$\text{definition of population mean: } \theta = \frac{\sum_{j=1}^{J} N_j \theta_j}{\sum_{j=1}^{J} N_j},$$

with corresponding estimate,

(2) $$\text{poststratified estimate: } \hat{\theta}^{\text{PS}} = \frac{\sum_{j=1}^{J} N_j \hat{\theta}_j}{\sum_{j=1}^{J} N_j}.$$

We use the general notation $\theta_j$ rather than $\overline{Y}_j$ to allow for immediate generalization to other estimands such as regression coefficients.

*Weighting.* When you look at sample survey data from a public-use dataset, the "survey weight" looks like a unit-level characteristic—just one more column in the data—and it is easy to think of it almost as a survey response, $w_i$. In this context it seems natural to use weighted averages of the form $\bar{y} = \sum_{i=1}^{n}(w_i y_i)/\sum_{i=1}^{n} w_i$.

But survey weights are *not* attributes of individual units—they are constructions based on an entire survey. Within any poststratification cell, all units have the same poststratification weight adjustment. (In theory, continuously-varying survey weights could arise from a survey with a continuous range of sampling probabilities. For example, one could imagine a survey of college-bound students where the probability of selection is a continuous function of background variables [e.g., $\Pr(\text{selection}) = \text{logit}^{-1}(a + b \cdot \text{SAT})$]. Or one could model nonresponse as a continuous function of predictors such as age and previous health status in a medical survey. These continuous weights do not come up much in the sorts of social surveys under consideration in this article, but they are interesting research directions that are potentially important in other areas of application.) We shall refer to *unit weights* $w_i$, $i = 1, \ldots, n$, and cell weights $W_j = n_j w_i$ for units $i$ within cell $j$,

$$\text{weighted average: } \bar{y} = \frac{\sum_{i=1}^{n} w_i y_i}{\sum_{i=1}^{n} w_i} \qquad (3)$$
$$= \frac{\sum_{j=1}^{J} W_j \bar{y}_j}{\sum_{j=1}^{J} W_j}.$$

Survey weights in general depend on the actual data collected as well as on the design of the survey. For example, consider the seven CBS polls conducted during the week before the 1988 Presidential election. These surveys had identical designs and targeted the same population. However, the weighting factor assigned to men (compared to a factor of 1 for women) varies from as low as 1.10 to 1.27 among the seven surveys. The different samples happened to contain different ratios of men to women and hence needed different adjustments.

*Weighting based on sampling probabilities.* A further complication is that survey weighting is commonly performed on some variables using inverse-sampling probabilities rather than poststratification. For example, in the Social Indicators Survey we assigned weights of 1/2, 1 and 2 for households with multiple phone lines, exactly one phone line and intermittent phone service, respectively. Unlike poststratification weights, these weighting factors are fixed and do not depend on the sample.

These inverse-probability weights are important in some survey designs and are sometimes portrayed as producing unbiased estimates, but this unbiasedness breaks down in the presence of nonresponse. For example, some telephone surveys give each respondent a weighting factor proportional to the number of adults in his or her household; this is an inverse-probability weight given that all households are equally likely to be selected (after correcting for the number of telephone lines) and the respondent is selected at random among the adults in the household. In practice, however, such weighting *overrepresents* persons in large households, presumably because it is easier to find someone at home from a household where more adults are living. Poststratification weights (which are roughly approximate to weighting by the square root of the number of adults in the households) give a better fit to the population (Gelman and Little, 1998).

In this article we shall assume that any factors associated with sampling weights have already been folded into the poststratification. For example, consider a survey that is poststratified into 16 categories (2 sexes × 2 ethnicity categories × 4 age ranges), and also has telephone weights of 1/2, 1 and 2. The three categories of telephone weights would then represent another dimension in the adjustment, thus giving a total of 48 categories. We recognize that treating this weighting as pure poststratification is an oversimplification; for one thing, sampling variances for poststratified estimates are generally different from those for fixed weights (see, e.g., Binder, 1983; Lu and Gelman, 2003).

### 1.3 Competing Methods of Estimation: Weighted Averages, Weighted Regression and Unweighted Regression Controlling for X

Many researchers have noted the challenge of using survey weights in regression models (as reviewed, e.g., by DuMouchel and Duncan, 1983; Kish, 1992; Pfeffermann, 1993). For the goal of estimating a population mean, it is standard to use the weighted average (3), but it is not so clear what to do in more complicated analyses. For example, when estimating a regression of $y$ on $z$, one recommended approach

TABLE 2

|  | respid | org | year | survey | y | state | edu | age | female | black | adults | phones | weight |
|---|---|---|---|---|---|---|---|---|---|---|---|---|---|
| 11352 | 6140 | cbsnyt | 7 | 9158 | NA | 7 | 3 | 1 | 1 | 0 | 2 | 1 | 923 |
| 11353 | 6141 | cbsnyt | 7 | 9158 | 1 | 39 | 4 | 2 | 1 | 0 | 2 | 1 | 558 |
| 11354 | 6142 | cbsnyt | 7 | 9158 | 0 | 31 | 2 | 4 | 1 | 0 | 1 | 1 | 448 |
| 11355 | 6143 | cbsnyt | 7 | 9158 | 0 | 7 | 3 | 1 | 1 | 0 | 2 | 1 | 923 |
| 11356 | 6144 | cbsnyt | 7 | 9158 | 1 | 33 | 2 | 2 | 1 | 0 | 1 | 1 | 403 |

Data from the first five respondents of a CBS pre-election poll. The weights are listed as just another survey variable, but they are actually constructed after the survey has been conducted, so as to match sample with known population information.

TABLE 3

| Opinion of NYC | True standard error | Different standard error estimates ||||
|---|---|---|---|---|---|
|  |  | assuming SRS | conditioning on weights | assuming inv-prob | design-based |
| Became a better place | 2.2% | 1.2% | 2.5% | 1.9% | 2.1% |
| Remained the same | 2.0% | 1.2% | 2.3% | 1.6% | 1.9% |
| Gotten worse | 2.0% | 1.2% | 2.4% | 1.7% | 2.0% |

From a simulation study: true standard error and four different standard error estimates for a question on the Social Indicators Survey. Ignoring the weighting or treating the weights as constant underestimates uncertainty, whereas uncertainty is overestimated by treating the weights as inverse probabilities. Accurate standard errors can be obtained using a jackknife-like procedure that explicitly takes account of the design of the weighting procedure. From Lu and Gelman (2003).

is to use weighted least squares, and another option is to perform unweighted regression of $y$ on $z$, also controlling for the variables $X$ that are used in the weighting.

Computing standard errors is not trivial for weighted estimates, whether means or regressions, because the weights themselves generally are random variables that depend on the data (Yung and Rao, 1996). In particular, correct classical standard errors cannot simply be obtained from the data and the weights; one also needs to know the procedure used to create the weights. Table 3 illustrates problems with some variance estimates that do not account for the weighting design. Similarly, with regressions, simple weighted regression procedures do not in general give correct standard errors.

### 1.4 The Crucial Role of Interactions

Consider a regression of $y$ on $z$, estimated in some way from a survey where inclusion probabilities depend on $X$. In general, $y$ can depend on both $X$ and $z$, in which case the appropriate way to estimate the regression of $y$ on $z$ is to regress $y$ on $X, z$ and then average over the population distribution of $X$. In general, estimating the regression of $y$ on $z$ requires estimation of the relation between $z$ and $X$ as well (Graubard and Korn, 2002). Because of the potential dependence of $z$ and $X$, it can be important to include interactions between these predictors in the model for $y$, even if the ultimate goal is simply to estimate the relation between $y$ and $z$.

In our survey adjustment framework, once a model includes interactions, poststratification is necessary in order to estimate population regression coefficients. For a simple example, suppose we are interested in the population regression of log earnings on height (in inches), using a survey that is adjusted to match the proportion of men and women in the population. The estimated regression (see Gelman and Hill, 2007) including the interaction is

$$y = \log(\text{earnings})$$
$$= 8.4 + 0.017 \cdot \texttt{height} - 0.079 \cdot \texttt{male}$$
$$+ 0.007 \cdot \texttt{height} \cdot \texttt{male} + \text{error}.$$

For any given height $z$, the expected value of log earnings is

$$\mathrm{E}(y|z) = 8.4 + 0.017z$$

$$
\begin{aligned}
(4) \quad & -0.079 \cdot \mathrm{E}(\mathtt{male}|\mathtt{height}=z) \\
& + 0.007 z \cdot \mathrm{E}(\mathtt{male}|\mathtt{height}=z).
\end{aligned}
$$

Here, $\mathrm{E}(\mathtt{male}|\mathtt{height}=z)$ would be estimated from the survey itself; most likely we would do this by fitting a linear regression (with Gaussian errors) of height given sex, and then simply using Bayes' rule, along with the population proportions of men and women, to compute the conditional probability.

The conditional expectation (4) is not, in general, a linear function of $z$. Thus, although we can define the population regression of log earnings on height—it is the result of fitting a simple linear regression of $y$ on $z$ to the entire population—it is not clear why it should be of any interest.

This difficulty in interpreting regression coefficients in the context of survey adjustments is one reason we have been careful in Section 1.1 to consider estimands that are simple comparisons of population averages. We illustrate with the goal of estimating the average difference in log earnings between whites and nonwhites; this is also a regression, but because the predictor $z$ is binary, it is defined unambiguously as a difference. Again, we suppose for simplicity that the survey is adjusted only for sex. The estimated regression fit, including the interaction, is

$$
\begin{aligned}
y &= \log(\text{earnings}) \\
&= 9.5 - 0.02 \cdot \mathtt{white} + 0.20 \cdot \mathtt{male} \\
&\quad + 0.41 \cdot \mathtt{white} \cdot \mathtt{male} + \text{error}.
\end{aligned}
$$

The population difference in log earnings is then

$$
\begin{aligned}
\mathrm{E}(y|\mathtt{white}&=1) - \mathrm{E}(y|\mathtt{white}=0) \\
&= -0.02 + 0.20 \cdot (\mathrm{E}(\mathtt{male}|\mathtt{white}=1) \\
&\qquad\qquad - \mathrm{E}(\mathtt{male}|\mathtt{white}=0)) \\
&\quad + 0.41 \cdot \mathrm{E}(\mathtt{male}|\mathtt{white}=1),
\end{aligned}
$$

and the factors $\mathrm{E}(\mathtt{male}|\mathtt{white}=0)$ and $\mathrm{E}(\mathtt{male}|\mathtt{white}=1)$ can be estimated from the data. More generally, this example illustrates that, once we fit an interaction model in a survey adjustment context, we cannot simply consider a single regression coefficient (in this case, for $\mathtt{white}$) but rather must also use the interacted terms in averaging over poststratification cells.

Our focus in this article is on the relation between the model for the survey response and the corresponding weighted-average estimate. The ultimate goal is to have a model-based procedure for constructing survey weights, or conversely to set up a framework for regression modeling that gives efficient and approximately unbiased estimates in a survey-adjustment context.

## 2. THE CHALLENGE

### 2.1 Estimating Simple Averages and Trends

We now return to the example of Table 1. The goal is to estimate $\overline{Y}^{2001} - \overline{Y}^{1999}$, the change in population average response between two waves of the Social Indicators Survey. This can be formulated as the coefficient $\beta_1$ in a regression of $y$ on time: $y = \beta_0 + \beta_1 z + \text{error}$, where the data from the two surveys are combined, and $z = 0$ and 1 for respondents of the 1999 and 2001 surveys, respectively.

A more general model is $y = \beta_0 + \beta_1 z + \beta_2 X + \text{error}$, where $\beta_2$ is a vector of coefficients for the variables $X$ used in the weighting. Now the quantity of interest is $\beta_1 + \beta_2(\overline{X}^{2001} - \overline{X}^{1999})$, to account for demographic changes between the two years. For New York City between 1999 and 2001, these demographic changes were minor, and so it is reasonable to simply consider $\beta_1$ to be the quantity of interest.

This brings us to the puzzle of Table 1. For each of three binary outcomes $y$, we compute the weighted mean for each year, $\bar{y}_w^{1999}$ and $\bar{y}_w^{2001}$, and two estimates of the change:

- Our first estimate is the simple difference, $\bar{y}_w^{2001} - \bar{y}_w^{1999}$, with standard error $\sqrt{\mathrm{var}(\bar{y}_w^{2001}) + \mathrm{var}(\bar{y}_w^{1999})}$, where the sampling variances are computed using the design of the weights (as in the rightmost column in Table 3).
- Our other estimate is obtained by linear regression. We combine the data from the two surveys into a single vector, $y = (y^{1999}, y^{2001})$, and create an associated indicator vector $z$ that equals 0 for the data from 1999 and 1 for the data from 2001. We fit a linear regression of $y$ on $z$, also controlling for the variables $X$ used in the weighting. (These $X$ variables are number of adults in the household, number of children in the family, number of telephone lines, marital status, and sex, age, ethnicity and education, and ethnicity $\times$ education for the head of household.) To estimate the change from 1999 to 2001, we use the coefficient of $z$, with standard error automatically coming from the (unweighted) regression.

As indicated in the third and fourth columns of Table 1, the regression coefficient and the change in weighted averages tend to have the same sign, but the two estimates sometimes differ quite a bit in magnitude. (Similar results are obtained if we work on the logit scale, as can be seen from the final two columns of the table.)

What should we believe? For this particular example, the direct analysis of weighted averages seems more believable to us, since we specifically created the weighting procedure for the goal of estimating these citywide averages. More generally, however, using weighted averages is awkward and we would prefer to use the more general techniques of regression and poststratification.

Where do we go from here? We would like an approach to statistical analysis of survey data that gives the right answers for simple averages and comparisons, and can be smoothly generalized to more complicated estimands.

### 2.2 Deep Poststratification

One of the difficulties of survey weighting is that the number of poststratification cells can quickly become large, even exceeding the number of respondents. This leads naturally to multilevel modeling to obtain stable estimates in all the poststratification cells, even those with zero or one respondent. Choices must then be made in the modeling of interactions.

For example, in our time-trend estimation problem, we could model $y = \beta_0 + \beta_1 z + \beta_2 X + \beta_3 X z +$ error, where $\beta_3$ is a vector of coefficients for the interaction of $X$ and time. We would then be interested in $\beta_1 + \beta_2(\overline{X}^{2001} - \overline{X}^{1999}) + \beta_3 \overline{X}^{2001}$ (as in the example at the end of Section 1.4). Where should the interaction modeling stop? A simulation study (Cook and Gelman, 2006) suggests that, in this example, efficient and approximately unbiased estimators are obtained by including interactions of the time indicator with all the survey adjustment factors; as a general approach, however, including all interactions can yield unstable estimates. The practical problem of adjusting for survey nonresponse leads to general questions of inference under multiway interactions, an issue that becomes even more relevant in small-area estimation.

Gelman and Carlin (2002) and Park, Gelman and Bafumi (2004) discuss the estimation of state-level opinions from national polls, using a hierarchical logistic regression with demographics and state effects, followed by poststratification on Census population totals for 64 demographic categories in each of the 50 states. The method worked well, but it is not clear how it would perform if the model included interactions of demographic and state effects.

## 3. USING REGRESSION MODELING TO CONNECT WEIGHTING AND POSTSTRATIFICATION

When cell means are estimated using certain linear regression models, poststratified estimates can be interpreted as weighted averages (Little, 1991, 1993). The idea is to work with the poststratified estimate (2)—an average over cell estimates $\hat{\theta}_j$, with the regression model providing the $\hat{\theta}_j$'s based on characteristics of the cells $j$. Under certain conditions, the poststratified estimate can be reinterpreted as a weighted average of the form (3), and then we can solve for the cell weights $W_j$ and the unit weights $w_i$.

### 3.1 Classical Models

*Full poststratification.* The simplest case is full poststratification of raw data, in which case the cell estimates are the cell means, $\hat{\theta}_j = \bar{y}_j$, and (2) becomes

$$\text{full poststratification: } \hat{\theta}^{\text{PS}} = \frac{\sum_{j=1}^{J} N_j \bar{y}_j}{\sum_{j=1}^{J} N_j},$$

which is equivalent to (3) with cell weights $W_j \propto N_j$ or unit weights $w_i \propto N_{j(i)}/n_{j(i)}$, where $j(i)$ is the poststratification cell to which unit $i$ belongs.

This estimate can also be viewed as a classical regression including indicators for all $J$ poststratification cells.

*No weighting.* The other extreme is no weighting, that is, unit weights $w_i = 1$ for all $i$, which is equivalent to poststratification if the cell estimates $\hat{\theta}_j$ are all equal to the sample mean $\bar{y}$, which in turn corresponds to classical regression including only a constant term.

*Classical regression on cell characteristics.* Intermediate cases of weighting can be obtained by regression models that include information about the poststratification cells without going to the extreme of fitting a least-squares predictor to each cell. For example, in the CBS/New York Times pre-election surveys, one could regress $y$ on indicators for sex, ethnicity, age, education and region, without necessarily including all their interactions.

Suppose the regression model is $y \sim \text{N}(X\beta, \sigma_y^2 I)$. We shall use $X$ to represent the $n \times k$ matrix of predictors in the data, and $X^{\text{pop}}$ to represent the $J \times k$ matrix of predictors for the $J$ poststratification cells. We also label the vector of poststratum populations as $N^{\text{pop}} = (N_1, \ldots, N_J)$, with a sum of $N = \sum_{j=1}^{J} N_j$.

The estimated vector of regression coefficients is then $\hat{\beta} = (X^t X)^{-1} X^t y$, and the estimated cell means are $X^{\text{pop}} \hat{\beta}$. The poststratified estimate of the population mean is then

classical regression:

$$\hat{\theta}^{\text{PS}} = \frac{1}{N} \sum_{j=1}^{J} N_j (X_j^{\text{pop}} \hat{\beta}) \tag{5}$$

$$= \frac{1}{N} (N^{\text{pop}})^t X^{\text{pop}} (X^t X)^{-1} X^t y, \tag{6}$$

which can be written as $\hat{\theta}^{\text{PS}} = \frac{1}{n} \sum_{i=1}^{n} w_i y_i$, with a vector of unit weights,

$$w = \left( \frac{n}{N} (N^{\text{pop}})^t X^{\text{pop}} (X^t X)^{-1} X^t \right)^t. \tag{7}$$

For convenience, we have renormalized these weights to sum to $n$ (see below). In (7), $w$ is a vector of length $n$ that takes on at most $J$ distinct values. The vector of $J$ possible unit weights (corresponding to units in each of the $J$ poststrata) is

$$w^{\text{pop}} = \left( \frac{n}{N} (N^{\text{pop}})^t X^{\text{pop}} (X^t X)^{-1} (X^{\text{pop}})^t \right)^t, \tag{8}$$

and the poststratified estimate can also be expressed as

$$\hat{\theta}^{\text{PS}} = \frac{1}{n} \sum_{j=1}^{J} w_j^{\text{pop}} \bar{y}_j.$$

The key result that makes the above computations possible—that allows $\hat{\theta}^{\text{PS}}$ to be interpreted as a weighted average of data—is that the derived unit weights $w$ in (7) sum to $n$. The identity $\sum_{j=1}^{n} w_j = n$ can be proved using matrix algebra but is more easily derived from an invariance in the classical regression model. With a least-squares regression, if a constant is added to all the data, that same constant will be added to the intercept, with the other coefficients not changing at all. Adding a constant to the intercept adds that same constant to $\hat{\theta}^{\text{PS}}$ in (5). We have thus established that adding a constant to each data point $y_i$ adds that same constant to $\hat{\theta}^{\text{PS}}$; thus, when $\hat{\theta}^{\text{PS}}$ is expressed as $\hat{\theta}^{\text{PS}} = \frac{1}{n} \sum_{i=1}^{n} w_i y_i$, these $w_i/n$'s must sum to 1.

The left panel of Figure 2 shows the unit weights obtained by fitting a sequence of classical regression models to the CBS/New York Times survey data. As more factors and interactions are included, the weights become more variable.

### 3.2 Hierarchical Models

We next consider the estimates that arise when applying the basic poststratification formula (2) when the cell means $\hat{\theta}_j$ are estimated using hierarchical models. As we shall see, we can formulate the resulting $\hat{\theta}^{\text{PS}}$ as a weighted average as in (3). In the classical estimates we have just considered, the equivalent weights $w_i$ depend on the structure of the model and the values of the predictors $X$ [see, e.g., (8)]. In contrast, with hierarchical models, we find that the $w_i$'s depend on the response variable $y$ being analyzed; for example, the vector of weights for the question on the respondent's health will be different than the vector of weights for the respondent's perception of the public schools. In our analysis we shall suppose that a particular response variable $y$ of interest has been selected (e.g., vote preference in the pre-election polls).

*Hierarchical regression.* The results in the previous section can be immediately generalized to multilevel regression models in which some of the coefficients are batches of indicator variables. We shall generalize the regression model to $y \sim N(X\beta, \Sigma_y)$ with a prior distribution on $\beta$ of the form $\beta \sim N(0, \Sigma_\beta)$. For simplicity, we assume independence of the components of $\beta$ in the prior distribution, conditional on hyperparameters for the variance components. [In practice, the covariance matrix $\Sigma_\beta$ would come from a fitted hierarchical model, and our analysis ignores uncertainty in the estimated hyperparameters. A fully Bayesian analysis would continue by averaging over the posterior distribution of $\Sigma_\beta$, which in turn would lead to a posterior distribution of equivalent weights; which might be summarized by a posterior mean, thus leading to posterior, or consensus, weights. Rao (2003) discusses this issue from a classical sampling-theory perspective.]

The prior precision matrix $\Sigma_\beta^{-1}$ is then diagonal, with zeroes for nonhierarchical regression coefficients (including the constant term in the regression). For example, consider a regression for the CBS/New York Times polls, with the following predictors:

- A constant term
- An indicator for sex (1 if female, 0 if male)

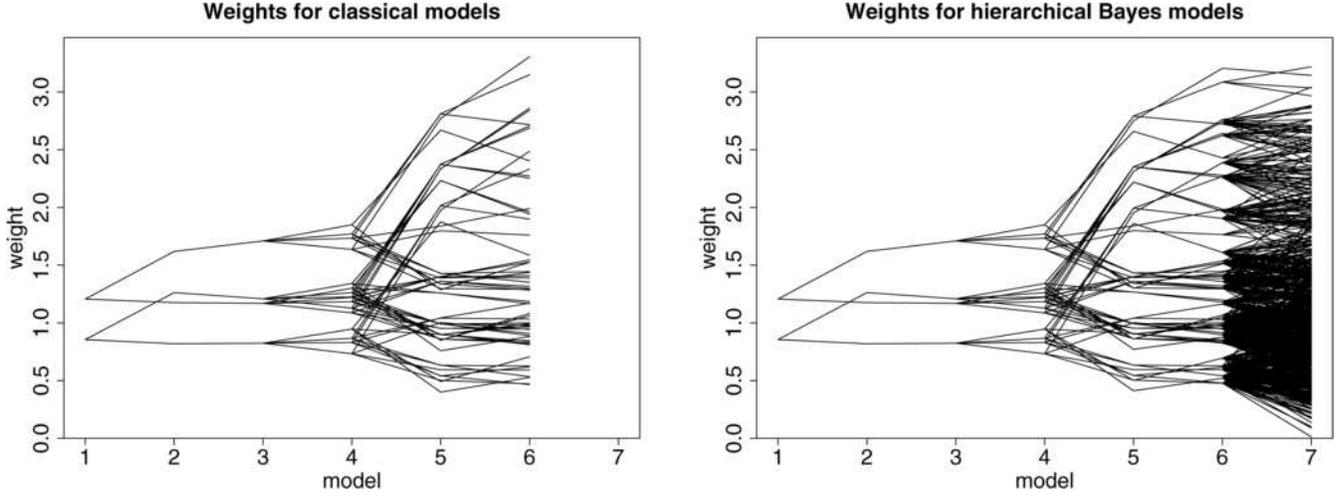

FIG. 2. *Equivalent unit weights $w_i$ for one of the CBS/New York Times surveys, based on a series of models fit first using classical regression and then using Bayesian hierarchical regression. The models are nested, controlling for (1) male/female, (2) also black/white, (3) also male/female × black/white, (4) also four age categories, (5) also four education categories, (6) also age × education and (7) also state indicators. Each model includes more factors and thus has more possible weights, which are renormalized to average to 1 for each model. For the Bayes models, the indicators for age, education, age × education and state are given independent batches of varying coefficients. For the classical weights, model (7) is not included because of collinearity.*

*The lines in each graph connect the weights for individual respondents, which are divided into successively more categories as predictors are added to the models.*

- An indicator for ethnicity (1 if black, 0 otherwise)
- Sex × ethnicity
- 4 indicators for age categories
- 4 indicators for education categories
- 16 age × education indicators.

The classical regression has $1+1+1+1+3+3+9=19$ predictors (avoiding collinearity by excluding the baseline age and education categories). The hierarchical regression has $1+1+1+1+4+4+16=28$ predictors, and its prior precision matrix has the form

$$\Sigma_\beta^{-1} = \text{Diag}(0,0,0,0, \sigma_{\text{age}}^{-2}, \sigma_{\text{age}}^{-2}, \sigma_{\text{age}}^{-2}, \sigma_{\text{age}}^{-2}, \sigma_{\text{edu}}^{-2},$$
$$\sigma_{\text{edu}}^{-2}, \sigma_{\text{edu}}^{-2}, \sigma_{\text{edu}}^{-2}, \sigma_{\text{age.edu}}^{-2}, \ldots, \sigma_{\text{age.edu}}^{-2}),$$

with the parameters $\sigma_{\text{age}}$, $\sigma_{\text{edu}}$ and $\sigma_{\text{age.edu}}$ estimated from data.

The estimated vector of regression coefficients is then $\hat{\beta} = (X^t \Sigma_y^{-1} X + \Sigma_\beta^{-1})^{-1} X^t \Sigma_y^{-1} y$ and expressions (6)–(8) become

Bayes poststratification:

$$\hat{\theta}^{\text{PS}} = \frac{1}{N}(N^{\text{pop}})^t X^{\text{pop}}$$
$$\times (X^t \Sigma_y^{-1} X + \Sigma_\beta^{-1})^{-1} X^t \Sigma_y^{-1} y,$$

(9) $\quad w = \left(\frac{n}{N}(N^{\text{pop}})^t X^{\text{pop}}\right.$
$$\left.\times (X^t \Sigma_y^{-1} X + \Sigma_\beta^{-1})^{-1} X^t \Sigma_y^{-1}\right)^t,$$
$$w^{\text{pop}} = \left(\frac{n}{N}(N^{\text{pop}})^t X^{\text{pop}}\right.$
$$\left.\times (X^t \Sigma_y^{-1} X + \Sigma_\beta^{-1})^{-1} (X^{\text{pop}})^t \Sigma_y^{-1}\right)^t.$$

Conditional on the variance parameters in $\Sigma_y$ and $\Sigma_{\text{beta}}$, then estimates from this model correspond to weighted averages.

The right panel of Figure 2 shows the unit weights obtained by fitting a sequence of Bayesian models to the CBS/New York Times poll. The first three models are actually identical to the classical (non-hierarchical) versions, since we assign noninformative uniform prior distributions to the coefficients for sex, ethnicity and their interactions. Models 4 and 5 are similar to the classical fits because age and education have only four categories, so there is little information available for partial pooling of these effects (see Gelman, 2006). The weights in model 6, with age × education interactions included, are smoothed somewhat compared to the corresponding classical model. Finally, introducing state effects



leads to a downweighting of some of respondents in states that happen to be overrepresented in the survey, and an upweighting for respondents in the undersampled states. There is no corresponding classical model here because the survey does not actually include data from all 50 states.

*Exchangeable normal model.* To understand these formulas better, we consider the special case of an exchangeable normal model for the $J$ cell means (see also Lazzeroni and Little, 1998; Elliott and Little, 2000). This model can be expressed in terms of the cell means,

$$\bar{y}_j \sim \mathrm{N}(\theta_j, \sigma^2/n_j),$$
$$\theta_j \sim \mathrm{N}(\mu, \sigma_\theta^2).$$

This is a special case of the hierarchical regression model discussed above, so we already know that the poststratified estimate, conditional on the (estimated) variance parameters $\sigma_y, \sigma_\theta$, can be expressed as a weighted average of the cell means, $\bar{y}_j$, or equivalently as a weighted average of the data points $y_i$.

In this simple example, however, we can gain some understanding by deriving algebraic expressions for the weights. Our goal is to express them in terms of the completely smoothed weights, $w_j = 1$, and the weights from full poststratification, $w_j = \frac{N_j/N}{n_j/n}$.

We start with the posterior means (conditional on the variance parameters) of the cell means. We write these as $\hat{\theta}_k, k = 1, \ldots, J$ (using $k$ as a subscript rather than $j$ because this results in more convenient notation later),

(10) $$\hat{\theta}_k = \frac{(n_k/\sigma_y^2)\bar{y}_k + (1/\sigma_\theta^2)\hat{\mu}}{n_k/\sigma_y^2 + 1/\sigma_\theta^2},$$

where

(11) $$\hat{\mu} = \frac{\sum_{k=1}^J \bar{y}_k/(\sigma_y^2/n_k + \sigma_\theta^2)}{\sum_{k=1}^J 1/(\sigma_y^2/n_k + \sigma_\theta^2)}.$$

We can combine (10) and (11) to express each $\hat{\theta}_j$ as a linear combination of the cell means $\bar{y}_k$,

$$\hat{\theta}_k = \sum_{j=1}^J c_{kj}\bar{y}_j.$$

After some algebra, we can write these coefficients as

$$c_{kj} = \begin{cases} \dfrac{\sigma_y^2}{n_k}A_k A_j/A, & \text{for } j \neq k, \\ \sigma_\theta^2 A_k + \dfrac{\sigma_y^2}{n_k}A_k^2, & \text{for } j = k, \end{cases}$$

where

$$A_k = \sum_{k=1}^J \frac{1}{\sigma_y^2/n_k + \sigma_\theta^2}.$$

The payoff now comes in computing the poststratified estimate,

$$\hat{\theta}^{\mathrm{PS}} = \sum_{k=1}^J N_k \hat{\theta}/N$$
$$= \sum_{k=1}^J \sum_{j=1}^J \frac{N_k}{N} c_{kj}\bar{y}_j,$$

equating this to $\sum_{j=1}^J W_j \bar{y}_j$ and thus deriving the cell weights,

$$W_j = \sum_{k=1}^J \frac{N_k}{N} c_{kj}$$
$$= A_j\left[\frac{N_j}{N}\sigma_\theta^2 + \sum_{k=1}^J \frac{N_k}{N}\frac{A_k}{A}\frac{\sigma_y^2}{n_k}\right].$$

The implicit unit weights are then $w_j^{\mathrm{pop}} = (n/n_j)W_j$, or

$$w_j^{\mathrm{pop}} = A_j\frac{n}{n_j}\left[\frac{N_j}{N}\sigma_\theta^2 + \frac{\sigma_y^2}{AN}\sum_{k=1}^J A_k\frac{N_k}{n_k}\right]$$

(12) $$= \frac{n}{\sigma_y^2 + n_j\sigma_\theta^2}$$
$$\times \left[\frac{N_j}{N}\sigma_\theta^2 + \frac{\sigma_y^2}{N}\frac{\sum_{k=1}^J N_k/(\sigma_y^2 + n_k\sigma_\theta^2)}{\sum_{k=1}^J n_k/(\sigma_y^2 + n_k\sigma_\theta^2)}\right].$$

The ratio of sums in (12) is a constant (given the fitted model) that does not depend on $j$. Let us approximate it by $N/n$ (which is appropriate if the sample proportions $n_k/N_k$ are independent of the group sizes $N_k$). Under this approximation, the unit weights can be written as

approximate $w_j^{\mathrm{pop}}$

(13) $$= \frac{n_j/\sigma_y^2}{n_j/\sigma_y^2 + 1/\sigma_\theta^2} \cdot \frac{N_j/N}{n_j/n}$$
$$+ \frac{1/\sigma_\theta^2}{n_j/\sigma_y^2 + 1/\sigma_\theta^2} \cdot 1,$$

which is a weighted average of the full poststratification unit weight, $\frac{N_k/N}{n_k/n}$, and the completely smoothed weight of 1. Hierarchical poststratification is thus approximately equivalent to a shrinkage of weights



by the same factors as in the shrinkage of the parameter estimates (10).

Thus, as with hierarchical regression models in general, the amount of shrinkage of the weights depends on the between- and within-stratum variance in the outcome of interest, $y$.

*Other hierarchical models.* Lazzeroni and Little (1998) and Elliott and Little (2000) discuss various hierarchical linear regression models, including combinations of the two models described above (i.e., a hierarchical regression with a cell-level variance component) and models with correlations between adjacent cell categories for ordered predictors.

Another natural generalization is to use logistic regression for binary inputs. Unfortunately, when we move away from linear regression, we abandon the translation invariance of the parameter estimates (i.e., the property that adding a constant to all the data affects only the constant term and none of the other regression coefficients). As a result, for logistic regression, the poststratified estimate $\hat{\theta}^{\text{PS}}$ is no longer a weighted average of the data, even after controlling for the variance parameters in the model. However, we suspect that the model could be linearized, yielding approximate weights.

### 3.3 Properties of the Model-Based Poststratified Estimates

*Standard errors.* The variance of the poststratified estimate, ignoring sampling variation in $X$, can be expressed using various formulas,

$$\text{var}(\hat{\theta}^{\text{PS}}) = \frac{1}{n^2} \sum_{i=1}^{n} w_i^2 \sigma_y^2 = \frac{1}{n} \sum_{j=1}^{J} (w_j^{\text{pop}})^2 n_j \sigma_y^2$$

$$= \frac{1}{nN} \sum_{j=1}^{J} w_j^{\text{pop}} N_j^{\text{pop}} \sigma_y^2.$$

Any of these equivalent expressions can be viewed as the posterior variance of $\theta$ given a noninformative prior distribution on the regression coefficients, and ignoring posterior uncertainty in $\sigma_y$ (Little, 1993).

*Dependence of implicit weights on the outcome variable.* Classical survey weights depend only on the $n_j$'s and the $N_j$'s, as well as the design matrix $X$ (used, e.g., to define the margins used in raking), but do not formally depend on $y$. (There is an informal dependence on $y$ in the sense that there is no urgency to weight on variables $X$ that do not help predict outcomes $y$ of interest.) Similarly, the implicit weights (7) obtained from a classical regression model depend only on $n$, $N$ and $X$, not on $y$.

However, the implicit weights (9) from hierarchical regression *do* depend on the data, implicitly, through the hyperparameters in $\Sigma_y$ and $\Sigma_\beta$, which are estimated from the data. Thus, the appropriate weights could differ for different survey responses.

## 4. WHERE TO GO NEXT

There are currently two standard approaches to adjusting for known differences between sample and population in survey data: weighting and regression modeling.

*Practical limitations of weighting.* The weighting approach has the advantage of giving simple estimates for population averages but has several disadvantages. First, it is not generally clear how to apply weights to more complicated estimands such as regression coefficients. There has been some work on weighted regression for surveys (e.g., DuMouchel and Duncan, 1983; Pfeffermann, 1993) but these procedures are not very flexible, which is one reason why the modeling approach is more popular for problems such as small-area estimation (Fay and Herriot, 1979). A second problem with weighted estimates is that standard errors are more difficult to evaluate (recall Table 3). Finally, weighting may be "dirty" but it is not always "quick": actually constructing the weighting for a survey is more difficult than you might think. Creating practical weights requires arbitrary choices about inclusion of weighting factors and interactions, pooling of weighting cells and truncation of weights. (For example, in the Social Indicators Survey, we decided to weight on some interactions and not others in order to control variability of the weights. While setting up the weighting procedure, we repeatedly compared weighted estimates to Census values for various outcomes that we thought could be "canaries in the coal mine" if the survey estimate did not fit the population. These "canary" variables included percentage of New York City residents who are U.S. citizens, the percent who own their own home and income quintiles.) The resulting vector of weights is in general a complicated and not-fully-specified function of data and prior knowledge. Subjective choices arise in virtually all statistical methods, of course, but good advice on creating weights tends to be much vaguer than for other methods in the statistical literature (see, e.g., Lohr, 1999).

*Practical limitations of modeling.* Regression modeling is easy to do—even hierarchical regression is becoming increasingly easy in Bugs, Stata and other software packages (see, e.g., Centre for Multilevel Modelling, 2005; Gelman and Hill, 2007)—but for analysis of survey data it has the disadvantage that, to combine with population information, the regression must theoretically condition on all the poststratification cells, which can lead to very complicated models—more complicated than we are comfortable with in current statistical practice—even in surveys of moderate size (see Section 2.2). When a model is too complicated, it becomes difficult to interpret or use the results, leading to awkward situations such as in Table 1, where we simply cannot trust the regression coefficients for time trends in the Social Indicators Survey.

It is a delicate point, because sometimes we do have confidence in regression coefficients, even with complicated hierarchical models with many parameters. For example, as discussed in Gelman and Carlin (2002) and Park, Gelman and Bafumi (2004), hierarchical regression combined with poststratification performs excellently at estimating state-level opinions from the national CBS/New York Times polls. So it is not just the number of parameters that is important, but rather some connection between the model and the quantities of interest, which is somehow more difficult to establish in the models whose results are shown in Table 1.

*Putting it together using hierarchical models and poststratification.* Our ideal procedure should be as easy to use as hierarchical modeling, with population information included using poststratification as in (1). The procedure should feature a smooth transition from classical weighting so that when different estimation methods give different results, it is possible to understand this difference as a result of interactions in the model (as discussed by Graubard and Korn, 2002).

How do we get there? One place to start is to focus on examples such as in Table 1 where different methods give different answers, and try to figure out which, if either, of the two estimates makes sense. A parallel approach is through simulation studies—for greater realism, these can often be constructed using subsamples of actual surveys—as well as theoretical studies of the bias and variance of poststratified estimates with moderate sample sizes. In addition, a full hierarchical modeling approach should be able to handle cluster sampling (which we have not considered in this article) simply as another grouping factor.

We would like a general modeling procedure that gives believable estimates for time trends and as a byproduct produces a good set of weights that can be used for simple estimands. Given the difficulties with current methods for weighting and modeling, we believe this approach is of both practical and theoretical interest.

## ACKNOWLEDGEMENTS

We thank Samantha Cook, John Carlin, Julian Teitler, Sandra Garcia, Danny Pfeffermann and several reviewers for helpful discussions and comments, and the National Science Foundation for financial support.